# Decoherence mitigation by real-time noise acquisition

## Authors

G. Braunbeck[1], M. Kaindl[1], A. M. Waeber[1], F. Reinhard[1,2,*]

[1] Walter Schottky Institut and Physik Department, Technische Universität München, Am Coulombwall 4, 85748 Garching, Germany

[2] Institut für Physik, Universität Rostock, Albert-Einstein-Str 23, 18059 Rostock, Germany

*friedemann.reinhard@uni-rostock.de

## Abstract

We present a scheme to neutralize the dephasing effect induced by classical noise on a qubit. The scheme builds upon the key idea that this kind of noise can be recorded by a classical device during the qubit evolution, and that its effect can be undone by a suitable control sequence that is conditioned on the measurement result. We specifically demonstrate this scheme on a nitrogen-vacancy (NV) center that strongly couples to current noise in a nearby conductor. By conditioning the readout observable on a measurement of the current, we recover the full qubit coherence and the qubit's intrinsic coherence time $T_2$. We demonstrate that this scheme provides a simple way to implement single-qubit gates with an infidelity of $10^{-2}$ even if they are driven by noisy sources, and we estimate that an infidelity of $10^{-5}$ could be reached with additional improvements. We anticipate this method to find widespread adoption in experiments using fast control pulses driven from strong currents, in particular in nanoscale magnetic resonance imaging, where control of peak currents of 100 mA with a bandwidth of 100 MHz is required.

## Introduction

Quantum sensing of weak signals often involves exposing a qubit to the signal during a long time[1–3], with noise being a limiting factor for maximum exposure time[4–6]. Manipulating the environment can decrease the noise arising from a few specific sources. For example, cooling to cryogenic temperatures reduces phononic noise[7], although at the cost of increased experimental complexity. Alternatively, the qubit can be controlled to reduce the effect of noise. Dynamical decoupling sequences tailor the spectral sensitivity of the qubit to a narrow frequency range and can drastically improve the coherence time[8–13]. Although they are not restricted to a specific decoherence source, they struggle with heavily fluctuating noise. Quantum error correction can artificially increase the decay constant of a qubit component storing the information by regularly checking the quantum state and actively correcting it[14–16], but analogous to classical memory correction, this requires redundant encoding, which can be infeasible for many systems.

Several state-of-the-art quantum science applications, such as quantum teleportation, rely on processing a quantum measurement during the experiment and feeding it forward to the next measurement to, for example, address the right qubit[17], apply a necessary transformation[18] or determine the appropriate readout basis[19]. In some cases, this is even possible while preventing full projection of the qubit by applying weak measurements for quantum feedback control[20].

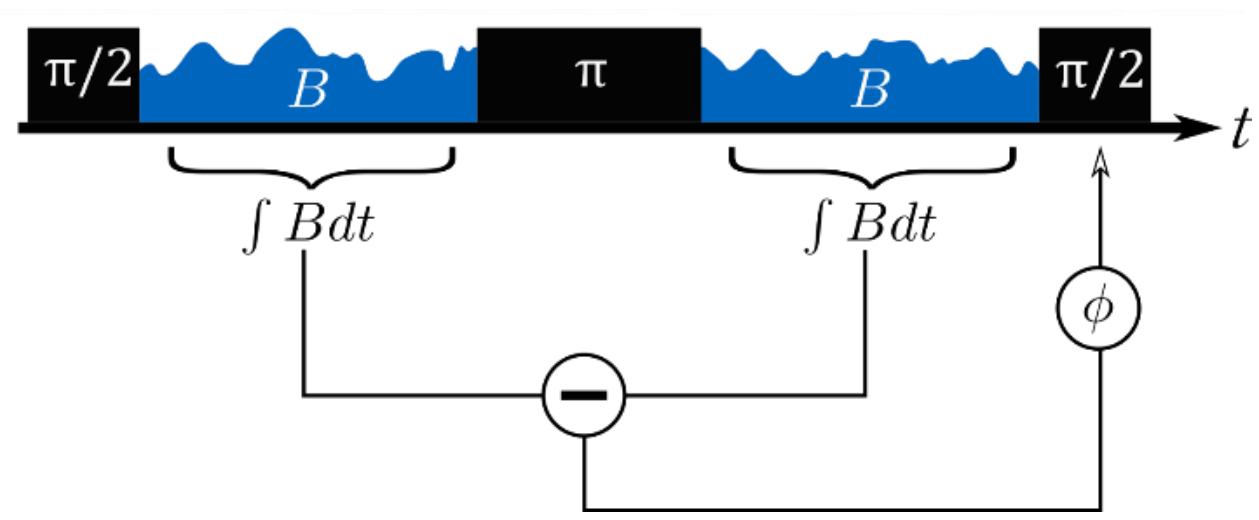

*Figure 1: Hahn Echo sequence with magnetic field noise during the free evolution times. If the noise is recorded, the phase of the final $\frac{\pi}{2}$-pulse can be adjusted accordingly to recover the full echo signal.*

In this work, we present quasi-feedforward (qff) decoupling, an approach that combines dynamical decoupling for constant and slow noise removal, with feedforward control, implemented via post-selection, for fast noise suppression. The basic principle is shown in Figure 1: We consider a qubit in a coherent superposition in the presence of noise during a free evolution period. The phase $\phi$ gained depends only on the integral of the noise during that period. If the exact course of noise is known throughout the free evolution, its effect is similar to a known and well-defined phase shift gate applied to the qubit and can be corrected. In order to achieve this, we feed forward for each repetition of the pulse sequence the recorded area of noisy current pulses that act as a classical noise source. If the phase of the final projection pulse is then shifted by the same amount as the spin during the free evolution period, the signal is fully restored. Note, that during the following experiments, instead of actively adapting the phase of the readout pulse, we chose to discard measurements where the phase of the readout pulse does not match the required phase for the measured current pulse. This will be discussed in detail during the experiment description. While this decision is based on limited hardware, it also demonstrates that our scheme works without the possibility of real-time control. Further, instead of adjusting the phase of the final projection pulse continuously, we adjusted the phase to the best fitting phase of four given values. Therefore, instead of fully restoring the signal, the signal was on average 90 % of the full signal.

Our method is similar in spirit to previous work to suppress dephasing by the nuclear bath of a quantum dot, where qubit control was adapted in real-time to a measurement of the nuclear Overhauser field [21]. However, in contrast to this prior work and quantum feedback in general, our scheme is limited to correcting the malicious effect of classical noise. The feedforward signal is based on a classical measurement of electric noise rather than a quantum measurement performed by the qubit itself.

Classical noise that can be measured simultaneously with the qubit evolution includes one main use case, which also motivated this work: suppressing decoherence from electronic noise in conductors near a qubit. In many emerging applications, solid-state qubits are addressed by conductors in close (sub-µm) proximity, in order to achieve strong Rabi frequencies for quantum control[22], strong coupling for Cavity-QED[23], or strong magnetic field gradients for nanoscale magnetic resonance imaging[24,25]. Strong coupling to these control signals comes at a price: electronic noise translates into strong decoherence. The most obvious solution – preventing noise altogether by low-noise control electronics – is often infeasible when strong control signals with large bandwidths are simultaneously required. This is particularly urging in two applications.

In electronic spin resonance (ESR) imaging – both at the macro- and microscale – gradient pulses need to have precisely controlled amplitudes and durations, because the relative error on these parameters directly defines the spatial resolution. At the same time, gradient currents have to be strong (100 mA-100 A). In contrast to nuclear magnetic resonance, they also have to be ramped on a fast (10 ns) timescale, because the response of spins is intrinsically fast and because the available coherence time is short[26] (µs). Classical low-noise current drivers like the Libbrecht-Hall design[27] cannot fulfill these needs, because of their limited closed-loop bandwidth (typically less than 1 MHz[28]). Faster control circuits are possible in theory (the gain-bandwidth-product of amplifiers can exceed 1GHz), but difficult in practice, for example because of strong sensitivity to capacitances in the signal path. In many ESR imaging experiments, closed-loop control has therefore been abandoned in favor of switching circuits that produce a defined current pulse by discharging a capacitor with a well-defined total charge[29,30]. These circuits, however, cannot be quickly switched between different current values and pulse durations, which limits the design of protocols.

In quantum control, many protocols involve simultaneous control of multiple qubit species. A prominent example is a dynamically decoupled controlled-NOT gate from a nuclear spin to a Nitrogen-Vacancy (NV) center qubit, which is a basic building block for both computation[31] and nuclear spin detection in sensing[3]. This operation requires a strong control pulse (a $\pi$ flip) on the nuclear spin, at a moment where the NV center is in a magnetically sensitive state. Because the nuclear magnetic moment is two to three orders of magnitude weaker than the electron magnetic moment of the NV center, crosstalk of the control pulse creates strong decoherence. At the same time, a low-noise design is difficult because of the simultaneous need for strong pulses and a high (>10 MHz) bandwidth.

While control of strong signals with simultaneously low noise and high bandwidth is challenging, a mere measurement of their fluctuations is much simpler. Fast acquisition hardware with bandwidths in excess of hundreds of MHz is becoming commonplace, and real-time control that can feed forward measurement results to subsequent control pulses is about to follow suit[32]. It thus appears timely to investigate adaptive quantum protocols which could reduce the requirements on control electronics in the applications mentioned above.

## Results

We demonstrate our concept in a geometry as displayed in Figure 2a (see supplementary material). A single NV center serves as the qubit system. It is placed in close (µm) proximity to a microfabricated wire in order to enable strong driving[22]. Currents in this wire couple strongly to the NV electron spin, with 1 mA typically translating into 1 G of magnetic field and 2.8 MHz of Zeeman shift. This enables the application of strong control pulses. For instance, a single-qubit phase shift gate can be driven with a Rabi frequency of up to $\Omega_z$ = 300 MHz by applying a current of 100 mA. However, the strong coupling also renders the qubit vulnerable to current noise, with 1 $mA_{rms}$ of current fluctuation shortening $T_2$ to only 50 ns. Active control of currents with this level of noise and bandwidth is technically challenging, which motivated our search for a scheme that replaces it with a mere measurement.

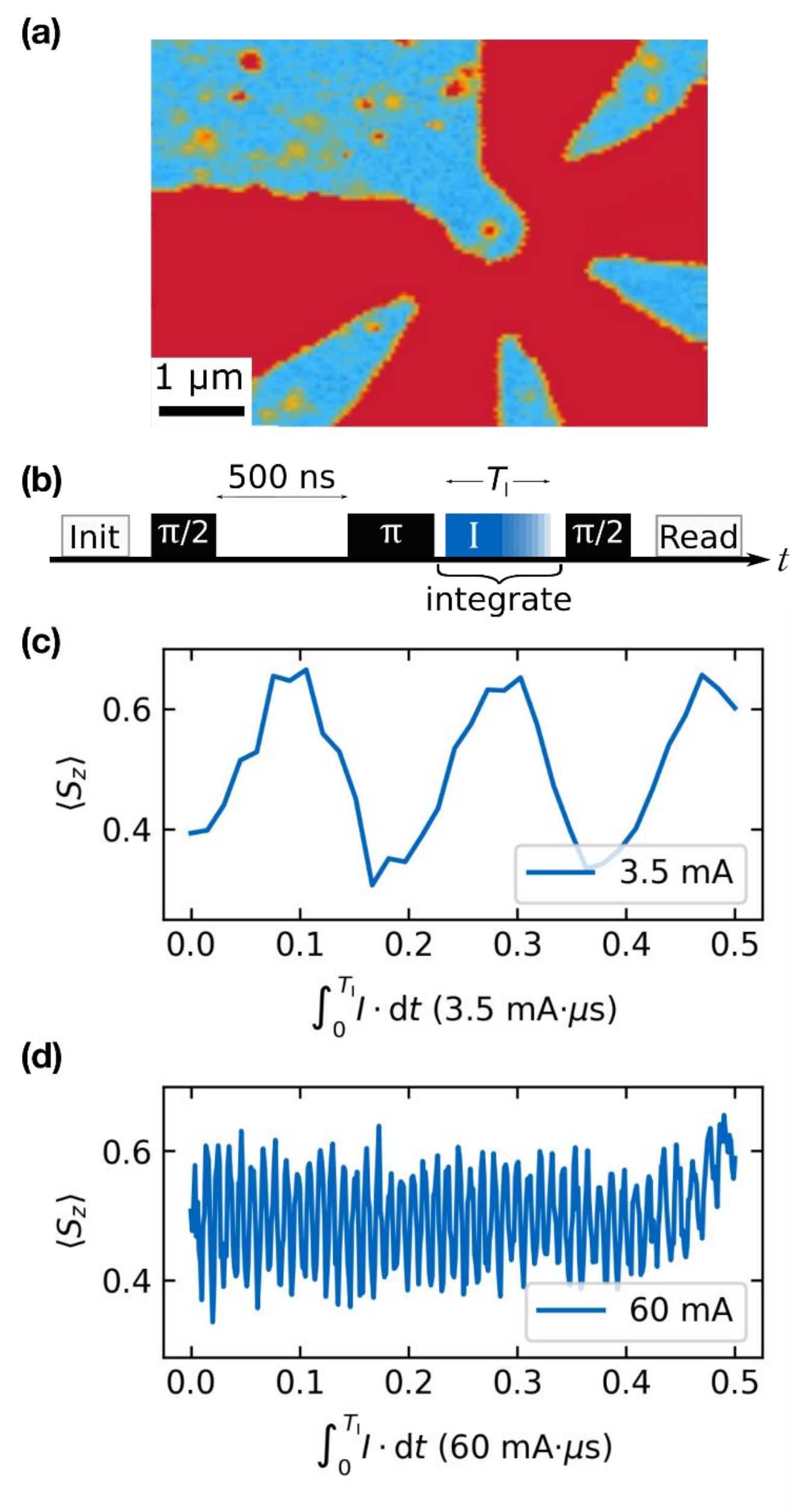

*Figure 2: a) Fluorescence scan of a diamond sample. A star shaped gold structure embraces a single nitrogen-vacancy center. Both qubit-control and noise field are applied via this structure. b) Measurement sequence. A Hahn echo sequence with fixed free evolution times of 500 ns is applied. Being on the order of $T_2$, this reduces the oscillation amplitude to 20 % of the full Rabi oscillation contrast. During the second free evolution time, a current pulse of fixed amplitude $I_0$ and varying duration $T_I$ is applied. c) + d) Oscillation of the spin qubit z-component depending on the current induced phase for currents with an amplitude of 3.5 mA (c) and 60 mA (d). The oscillation frequency scales with the amplitude.*

We first apply this paradigm to a challenge of quantum control: to apply a long phase shift gate under the influence of current fluctuations, as it is required in Fourier imaging [24]. Unknown fluctuations severely limit the maximum phase that can be gathered coherently. In order to achieve on the order of 100 full phase rotations, we use current measurements and quasi-feedforward decoupling to protect the qubit against decoherence from current fluctuations (Figure 2).

The experimental sequence is displayed in Figure 2b. We use a slightly altered Hahn echo sequence, where we insert a current pulse during the second free evolution time. Another NV-specific necessity is the projection of the phase onto the NV's z-axis with an additional $\frac{\pi}{2}$-pulse before readout. The microwave control pulses are provided by a source (Rohde&Schwarz SMIQ06b) that is split by a 90°-splitter (Mini-Circuits ZAPDQ-4+), yielding an in-phase and a quadrature component. These are switched (Mini-Circuits ZASWA-2-50DR+) individually to enable the use of $x$-phase and $y$-phase pulses for later experiments. Both components are then combined (Mini-Circuits ZX10-2-442+) into one line and amplified (Mini-Circuits ZHL-16W-43+). The current comes from a direct current source (KORAD KA3005P) that is controlled by a high-side switch (iC-Haus iC-HGP), combined with the microwave signal in a bias tee (Taylor BT-A1080-3), and finally sent towards the gold structure around the NV center.

The current's magnetic field at the NV position leads to a time dependent detuning

$$\Delta(t) = \Delta_0 \times I(t) \tag{1}$$

with constant $\Delta_0$, and thus increases the NV phase by

$$\phi = \int_0^{T_\mathrm{I}} \Delta(t)\ \mathrm{d}t\ = \Delta_0 \int_0^{T_\mathrm{I}} I(t)\ \mathrm{d}t \tag{2}$$

inducing an oscillation of the NV spin with increasing integrated current (Figures 2c and 2d). The oscillations have a contrast of less than unity because of intrinsic decoherence during the Hahn echo. For each experimental repetition, we measure and integrate the respective current before correlating it to the qubit readout result. A differential probe (Yokogawa 701920, see supplementary material) measures the current flow by monitoring the voltage drop over a 50 Ω resistor that is in series with the current application structure. An oscilloscope (Spectrum M4i.4451-x8) with 500 MS/s and a 14-bit digital resolution is recording the voltage for every repetition, allowing us to correlate each readout result with its preceding current pulse. Specifically, NV photons collected during the readout pulse are registered and binned by their measured pulse area $\int_0^{T_I} I(t)\ \mathrm{d}t$ . In this way, photons will contribute to the signal at the actual value of pulse area, not the nominal setpoint chosen by pulse duration and/or current amplitude.

The recorded oscillations in Figure 2 both show a decay constant $T_2$ similar to the intrinsic constant $T_2 = (1.4 \pm 0.1)\ \mu\mathrm{s}$. In other words, the original signal drops to 1/e over $130 \pm 9$ full oscillations. If we interpret the current pulse as a $260\ \pi$ phase gate, this corresponds to a gate infidelity per π-pulse of $8 \times 10^{-3}$ (see supplementary material). Note that this number only refers to the phase imprinted by the current pulse. The fidelity of the ($\pi/2$ and $\pi$) microwave control pulses is likely lower. This measurement also provides a calibration of $\Delta_0 = (9.48 \pm 0.08)\,\frac{\mathrm{rad}}{\mathrm{mA}\cdot\mu\mathrm{s}}$.

We now focus on a different challenge: to recover an echo envelope in the presence of strong fluctuations of the current by applying quasi-feedforward decoupling (Figure 3). This scenario is representative for all measurements that involve strong fields or control pulses, such as imaging experiments that require a strong magnetic field gradient[24]. As before, we perform an echo decay measurement, which is perturbed by a current pulse during the second free evolution period. The current pulses are of random duration ranging from 0 ns to $\tau$ with an amplitude of 3.50 ± 0.02 mA and are measured in each experimental repetition. The measured current is converted into a phase by the previously calibrated $\Delta_0$ and fed forward to the phase of the final $\frac{\pi}{2}$ -pulse in order to maximize the echo signal. This step could either be performed by real-time electronic processing or can be emulated by post-selection, which is the approach we choose for this study. Specifically, we alternate the phase of the final $\frac{\pi}{2}$-pulse between subsequent repetitions (Figure 3a), implementing alternating readout of the observables $\sigma_x$ and $\sigma_y$. We then discard those measurements that do not match the feedforward phase computed above (Figure 3b).

For a better understanding, we use the Bloch sphere visualization for the gathered phase during the second free evolution period, so that the phase corresponds to the azimuthal angle $\phi$. In more detail, we discretize the gathered phase into four quadrants (Figure 3b), corresponding to maximum positive *z*-component after a $\frac{\pi}{2}$-pulse with x- or y-phase (*x+*, illustrated in Figure 3c, and *y+*), and maximum negative *z*-component after a $\frac{\pi}{2}$-pulse with x/y-phase (*x-/y-*). The phase of the final $\frac{\pi}{2}$-pulse is alternating between x and y and we only keep the measurement if the combination of gathered qubit phase and the phase of the final $\frac{\pi}{2}$-pulse maximizes the absolute value of the qubit's final *z*-component and thus the signal contrast. The dark blue curve in Figure 3d shows the measurement outcome in the presence of the current noise of the random-duration current pulses without post-selection. The result is an exponential decay with $T_2 = (55 \pm 13)$ ns, which is much shorter than the measured decay of the echo in the absence of any current pulse (Figure 3d, light blue curve), where the decay constant is $T_2 = (389 \pm 133)$ ns. The red and green curves are the post-selected results of our quasi-feedforward decoupling method for x- and y-projection pulses, labeled $\mathrm{qff}_x$ and $\mathrm{qff}_y$. The method corrects the perturbing effect of the random-duration current pulses, increasing the coherence time sevenfold to $T_2 = (366 \pm 133)$ ns, reaching the value of the unperturbed measurement within the measurement accuracy.

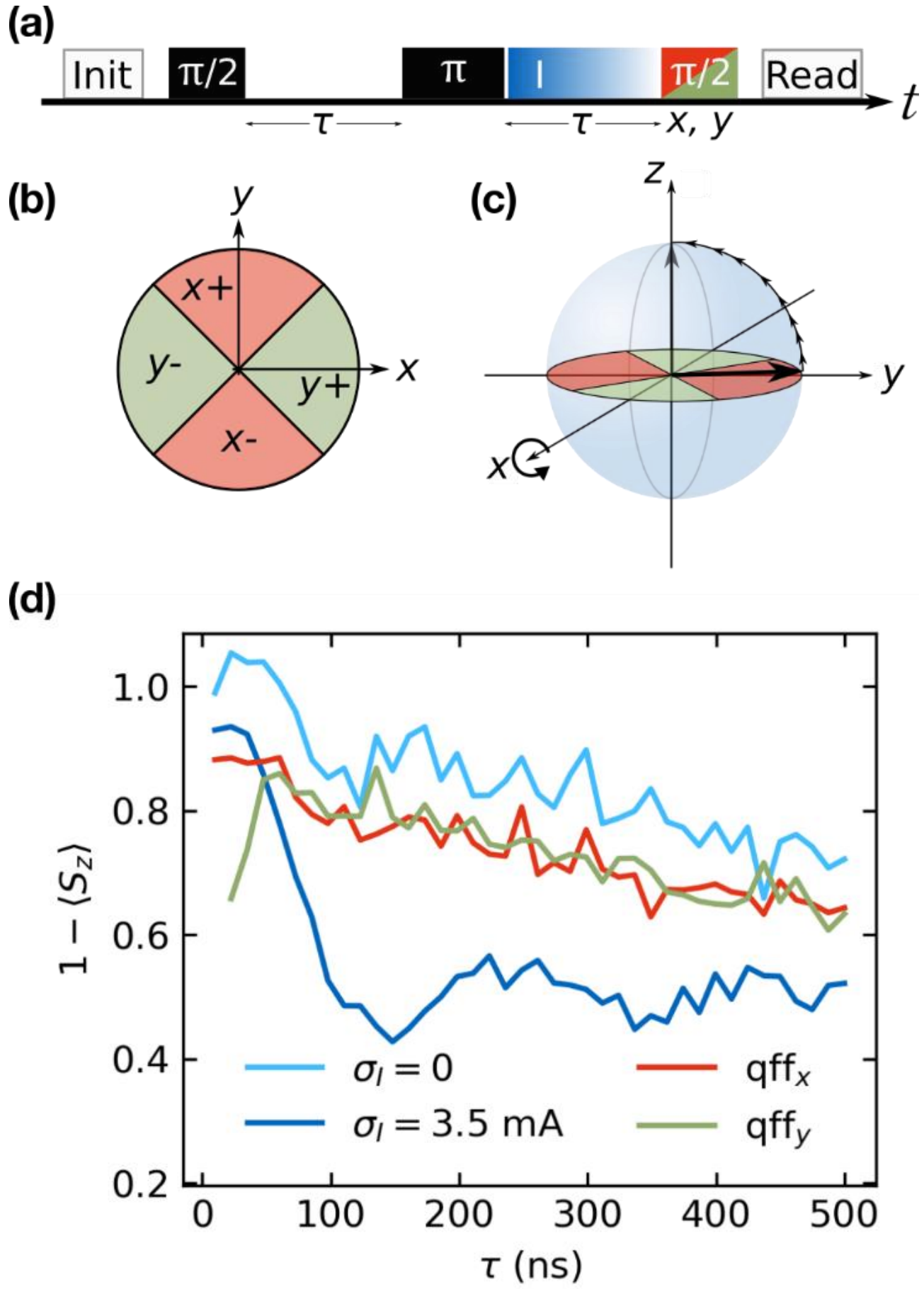

*Figure 3: a) A Hahn echo sequence with variable free evolution time $\tau$. During the second free evolution time, a current pulse of random length is applied. The phase of the final $\frac{\pi}{2}$-pulse before readout is alternated between x and y. b) Top view of a Bloch sphere equatorial plane. We divide the unit circle into four quadrants, each representing a potential projection phase to maximize the signal amplitude $1 - \langle S_z \rangle$ for a given qubit state at $2\tau$. A state in a red quadrant requires a final $\frac{\pi}{2}$-pulse with x-phase, a green one a $\frac{\pi}{2}$-pulse with y-phase to maximize the signal. c) Bloch sphere with the quadrants from b) indicated. The bold black arrow depicts a possible qubit state before the last $\frac{\pi}{2}$-pulse in a red quadrant. A $\frac{\pi}{2}$-pulse with x-phase invokes the indicated rotation, eventually bringing the qubit state parallel to the z-axis. A pulse with y-phase would invoke a rotation around the y-axis, which in this case would yield barely any contrast. d) Measurement results of the sequence shown in a). The light blue graph is the result in absence of noise, decaying with constant $T_2 = (389 \pm 133)\ ns$. The dark blue graph shows the result in the presence of noise without any post-selection, leading to the short decaying constant $T_2 = (55 \pm 13)\ ns$. The red (green) graph, derived from the measured data by quasi-feedforward decoupling, neglects all results but those where an x-phase (y-phase) projection pulse was applied on the qubit system in one of the red (green) quadrants of b).*

The effective noise reduction of our quasi-feedforward scheme is of special interest for quantum sensing applications (see discussion section). During the free evolution time of a Hahn echo or related sequence, the NV center becomes susceptible to magnetic fields[33], turning the sequence effectively into a sensing protocol. The measured field is stored in the phase $\phi$. In a realistic experiment, the field consists of the desired signal $B_{\mathrm{sig}}(t)$ and measurement corrupting noise $B_{\mathrm{noise}}(t)$. When the noise is recordable, for example when it originates from a current, the gathered phase of a measurement can be written analogously to equation (2) as

$$\phi = \gamma \int_0^\tau \left[B_{\text{sig}}(t) + B_{\text{noise}}(t)\right] \, \mathrm{d}t \overset{(1)}{=} \gamma \int_0^\tau B_{\text{sig}}(t) \, \mathrm{d}t \ + \Delta_0 \int_0^\tau I(t) \, \mathrm{d}t \ , \quad (3)$$

with the gyromagnetic ratio $\gamma$. The latter part can be removed by quasi-feedforward correction as explained above, leaving only the phase gathered by the desired signal even under the influence of noise.

On the one hand, inaccuracy per measurement is at maximum half a sector width. So accuracy is generally maximized by choosing sectors as small as possible. On the other hand, this comes at the cost of discarding most measurements. In order to quantify this tradeoff and optimize the sector number, we estimate the sensitivity of such a sensing protocol in the limit of strong electric noise that spreads the (uncorrected) phase uniformly across the full equator of the Bloch sphere ($\sigma_\phi \gg 2\pi$). With quasi-feedforward correction, this phase spread will be reduced to the width of a single phase sector. The corrected phase varies within $\pm\sigma_\phi$ where $\sigma_\phi = \frac{\pi}{N_\phi}$, with $N_\phi = 2{,}4{,}6$ … being the number of sectors. The expectation value of the measured $z$-component of the spin, which corresponds to the readout fidelity, is then

$$\langle S_z \rangle = \int_{-\sigma_\phi}^{\sigma_\phi} \cos(\theta) \mathrm{d}\theta / 2\sigma_\phi = \mathrm{sinc}(\sigma_\phi) = \mathrm{sinc}(\frac{\pi}{N_\phi}). \quad (4)$$

With increasing $N_\phi$, the readout fidelity asymptotically approaches one, however, the number of discarded measurements is growing as well. Only two opposing sectors are usable for a certain gained phase. Consequently, the fraction of usable measurements $\frac{2}{N_\phi}$ approaches zero with increasing $N_\phi$.

The resulting sensitivity $\sigma_B$ of a Hahn echo measurement is therefore

$$\frac{\sigma_B}{\sigma_{B,0}} = \frac{\sqrt{N_\phi/2}}{\mathrm{sinc}(\frac{\pi}{N_\phi})} \quad (5)$$

where $\sigma_{B,0} = \hbar/(g\mu_B\sqrt{T_2})$ is the standard quantum limit[34], the numerator $\sqrt{N_\phi/2}$ encodes the overhead induced by discarded measurements, and the denominator $\mathrm{sinc}(\frac{\pi}{N_\phi})$ denotes the loss of spin contrast due to the remaining phase uncertainty (Figure 4). Considering, that $N_\phi$ must be an even integer, the ideal number of sectors is finally determined to be either $N_\phi = 2$ or $N_\phi = 4$, where sensitivity is degraded by a factor 1.57, corresponding to a 2.47 longer acquisition time. Interestingly, this indicates, that all experiments shown would have performed just as well with only two sectors instead of four.

Although the readout fidelity is reduced to $\langle S_z \rangle$ = 90 % for $N_\phi = 4$, this is a static readout handicap that does not scale with the amount of applied qubit gates, exemplarily π-pulses. In general, the gate fidelity can thus be much higher, as the shown experiments prove with gate infidelities on the order of $10^{-2}$ (see supplementary material).

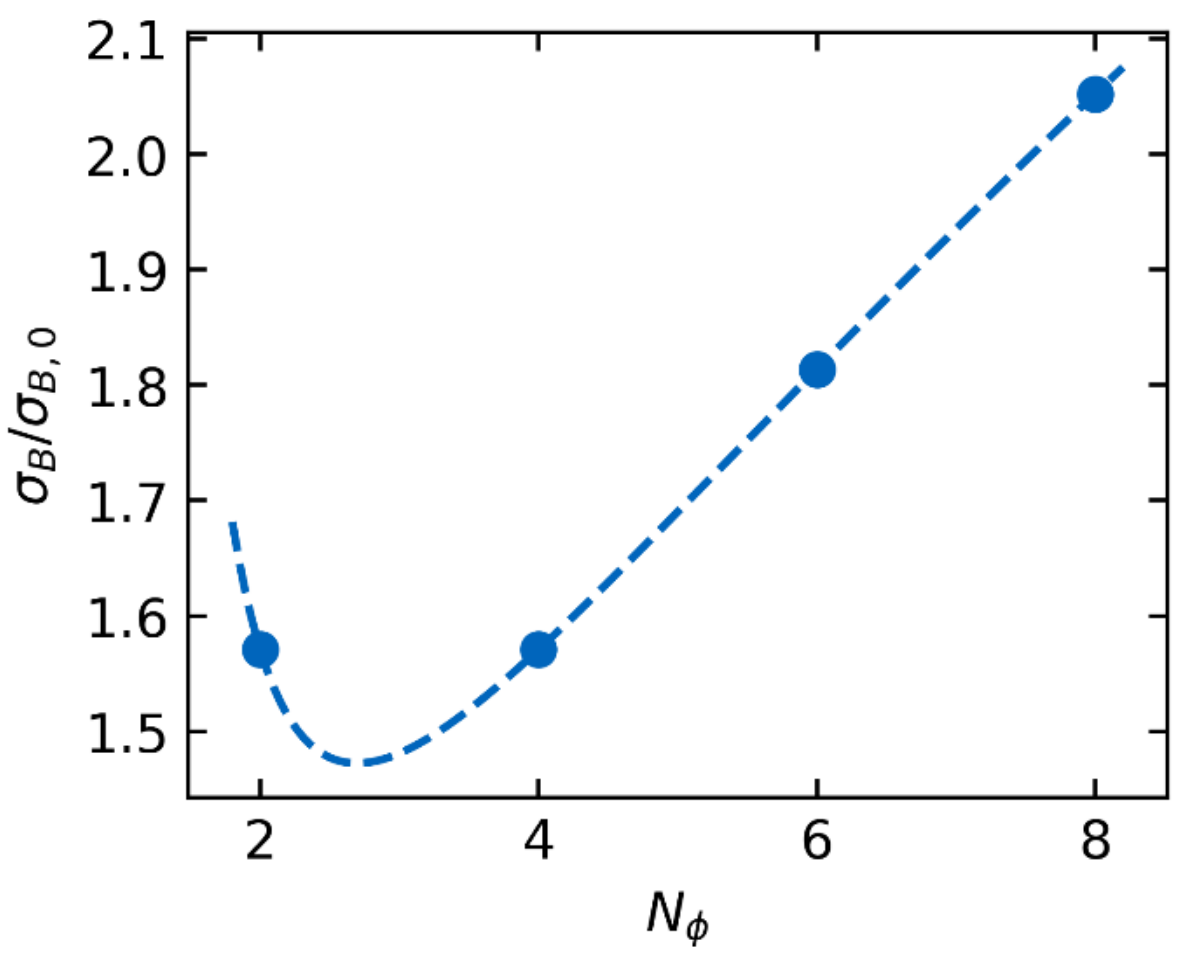

*Figure 4: Sensitivity scaling. The dashed line shows the simulated scaling of the sensitivity $\sigma_B$ with the number of sectors $N_\phi$. The graph is plotted for a theoretically continuous number of sectors. The theoretical minimum would be achieved for $N_\phi = 2.7$. However, for the presented quasi-feedforward scheme, only even integers are allowed, marked by blue dots. With this restriction, the optimum number of sectors is $N_\phi = 2$ or $N_\phi = 4$.*

We now extend quasi-feedforward decoupling to a scheme, which does not require pre-calibration of the qubit response and instead learns the value of $\Delta_0$ on the fly. We record and save the free evolution duration $\tau$, the integrated current $\int I\mathrm{d}t$, the projection pulse phase, still only allowing *x* and *y,* and the readout result for every repetition (Figure 5). Finally, we correct for the decoherence caused by the current noise in post-processing.

The triangular shape of the 2d-plot in the top left panel of Figure 5 is associated with the boundary condition $T_I \leq \tau$, which limits the integrated current to low values for short $\tau$. Plotting oscillation amplitudes against $\tau$, shown below the 2d-plot, provides the expected echo decay. The projection along the $\int I\mathrm{d}t$-axis, shown to the right of the 2d-plot, exhibits a phase oscillation comparable to those in Figures 2b) and c). From this data, we can extract the calibration analogously to before and apply the same post-selected quasi-feedforward decoupling correction. We note that the coupling to the current and the oscillation frequency differ from the data of Figure 2 as this experiment was conducted on a different NV center.

This self-learning approach is faster than the pre-calibration approach of Figures 2 and 3, since every measurement contributes simultaneously to both the echo envelope and the calibration of the qubit response. This also implies that the accuracy of the post-measurement calibration is the same as for the previous method: N measurements provide N data points for the calibration curve in both cases. Further, instead of the simple quadrant calibration introduced in Figure 3b), the post-measurement calibration allows for example to fit a sine curve for every value of $\tau$. This performs better especially for boundary cases, where $\phi$ lies close to the quadrant boundaries in Figure 3b). In future work, machine learning and Bayesian techniques could further improve efficiency[21,35].

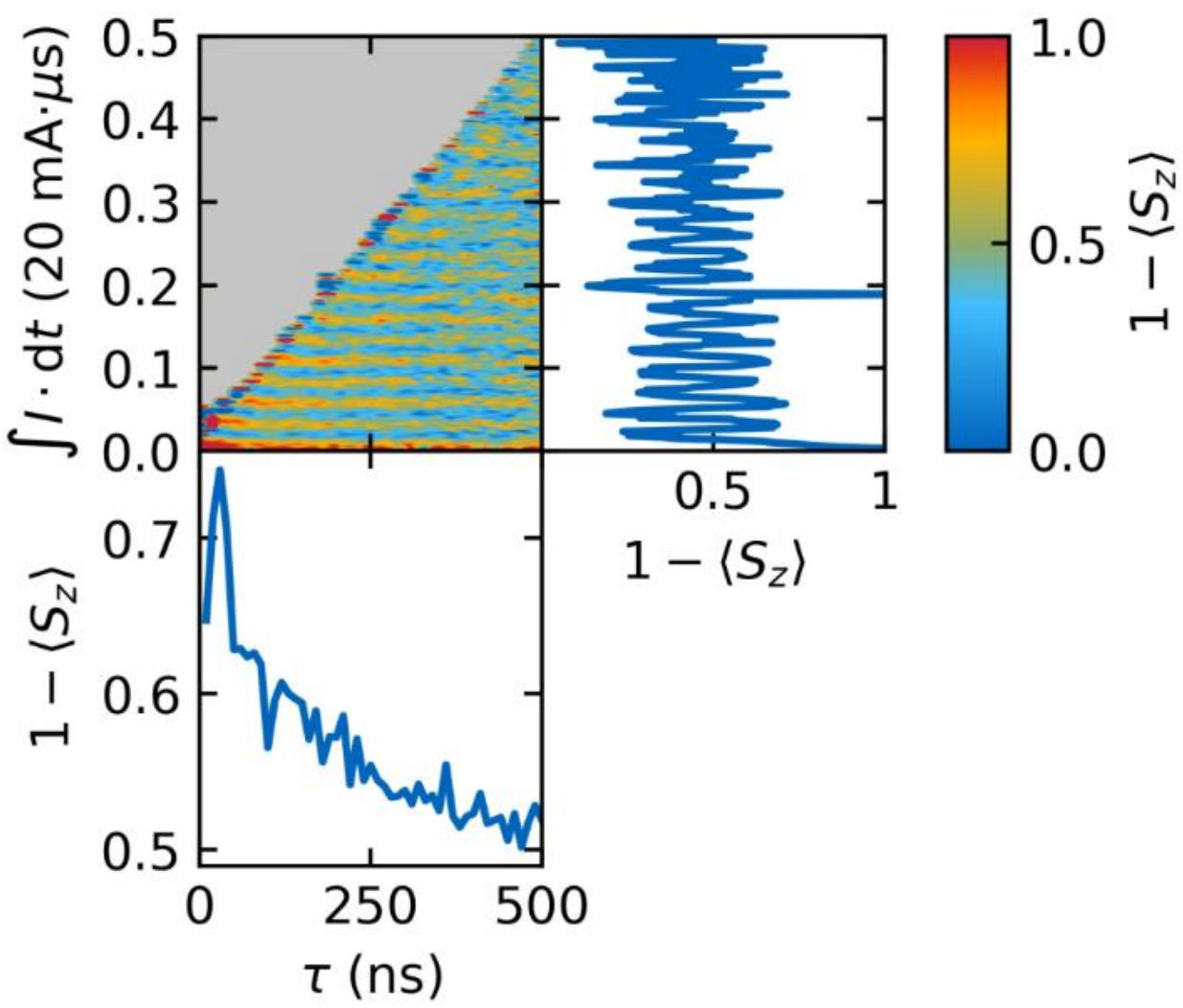

*Figure 5: Post-measurement calibration. The measurement result for different $\tau$ and $T_I$ is shown in the 2d-plot of the top left panel. The projection of the data onto the integrated-current axis, shown in the top right panel, exhibits a phase oscillation, corresponding to the one used for pre-calibration. It is used for a post-measurement calibration. The oscillation amplitude for each $\tau$, shown in the bottom left panel, shows the expected echo decay.*

We now turn to a benchmark of our method. Similar to the measurement in Figure 2, we fix the free evolution period of an echo to 32 µs and sweep the duration of a current pulse with nominally constant amplitude. We compare the resulting oscillation amplitude over time, once without any correction and once corrected for current fluctuations as described above. The echo decays noticeably faster without correction. Using the measured amplitude and pulse duration for correction, we can extend the decay constant $T_2$ from $(12.7 \pm 0.5)$ µs to $(17.6 \pm 0.7)$ µs in our experiment (Figure 6a). Figures 6b – 6d show the recorded oscillation, zoomed in for certain time periods. Both curves behave the same for very short (Figure 6b) and very long current (Figure 6d) pulses. However, there is a significant difference at intermediate pulse lengths (Figure 6c). With quasi-feedforward decoupling, there is still a significant oscillation visible, while the uncorrected measurement has already decayed to pure noise. Given an oscillation frequency of $\Delta\omega_z = 10$ MHz, our technique enables us to perform 38% more full oscillations than the classical noise of our current source allows.

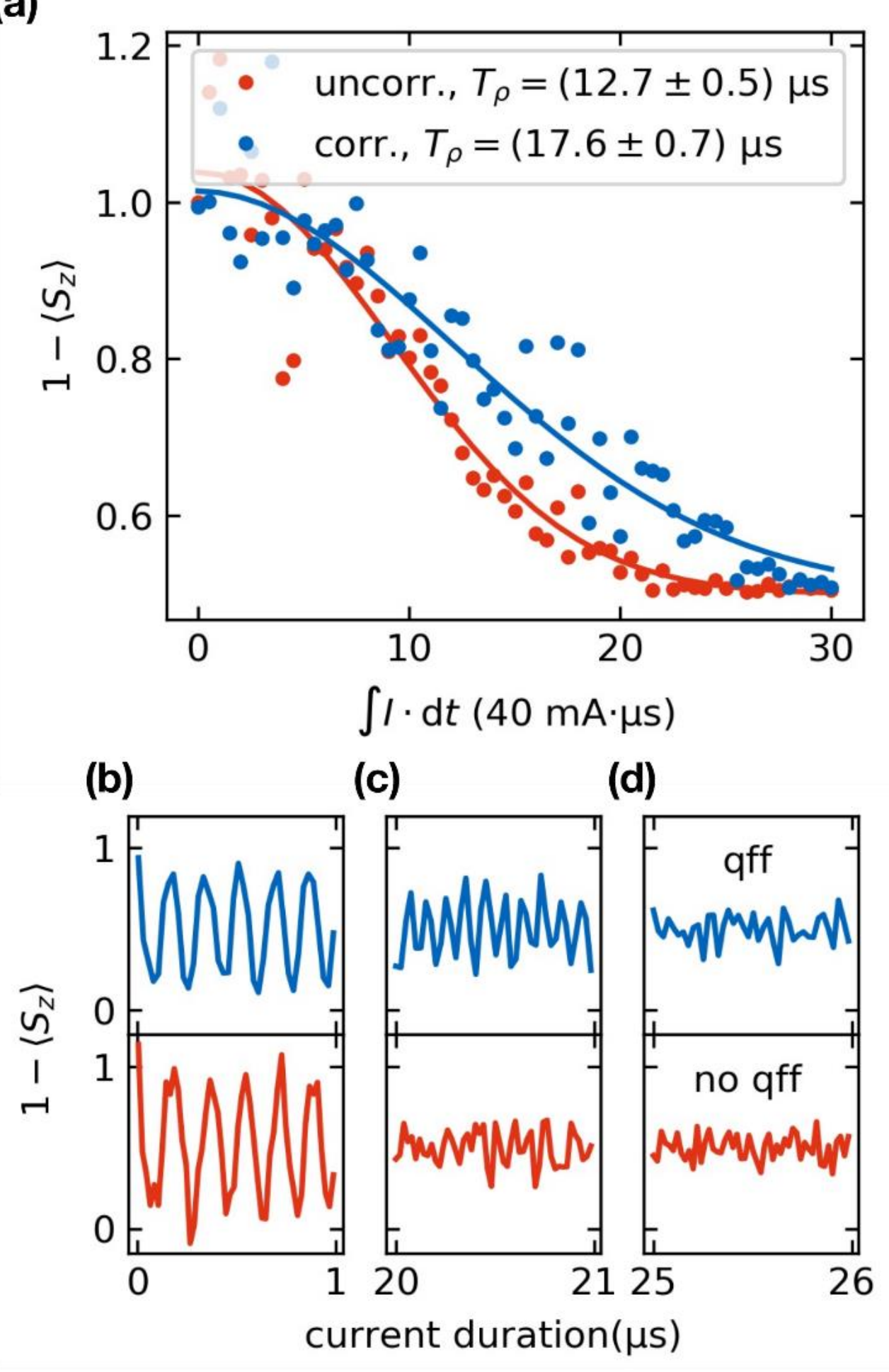

*Figure 6: Echo benchmark. The duration of a current pulse is swept during the second free evolution period of an echo with fixed free evolution time of* $32\ \mu s$ *(compare Figure 2b). We compare the decay of the resulting oscillation graph with (blue) and without correction (red). The current pulse is nominally set to 40 mA and swept from 5 ns to 30 µs. a) The correction extends the decay constant* $T_2$ *from* $(12.7 \pm 0.5)$ *µs to* $(17.6 \pm 0.7)$ *µs. b), c) and d) Exemplary comparisons of the oscillations with and without correction. The most dramatic difference can be seen in c), where our technique clearly recovers an oscillation that would otherwise be lost.*

The discrepancy between the correction performance in Figures 3 and 6 is explained by significant differences in the samples. The sample used in Figure 3 is a Ib diamond and the wire structure around the NV center is grown directly onto the diamond. The intrinsic $T_2$ is therefore low but the current flow through the structure is rather stable. The task in Figure 3 is thus restoring the decay induced from a few oscillations over less than one micro second.

The sample used in Figure 6 is a IIa diamond and the wire structure is grown on a glass plate, that is bonded with translucent polymer onto the diamond. The intrinsic $T_2$ is therefore high but the current flow is likely fluctuating much stronger by thermal disposition. The task in Figure 6 is thus restoring the decay induced by hundreds of oscillations over tens of micro seconds.

## Discussion

Feed-forward decoupling succeeds in mitigating noise from fluctuating control currents in all three different schemes tested in this study. Quantitatively, the improvement afforded is weaker than expected. For all our experiments, the gate infidelity is approximately $10^{-2}$. In an imaging experiment, this would limit localization precision to approximately 25 nm (see supplementary material). This infidelity is much higher than the ultimate limit, which is set by the digitization noise of the oscilloscope. With a 14-bit digitization accuracy and typically 25 measurements per $\pi$ phase shift gate, future experiments reaching this limit should be able to reach a gate infidelity as low as $10^{-5}$.

The most plausible interpretation of this discrepancy is that the method is successful in eliminating current noise, but that current noise is not the dominating source of dephasing in our present device. There are several other effects that could provide such a source. One possible candidate is heat expansion of the conductors. As the conductors have a cross-section of ~1 µm², a fluctuating expansion of 2.5% would displace the current by 25 nm and explain the measured infidelity. However, this would imply a temperature change of more than 1000 K, which seems unlikely. Another problem might be the granularity of the electroplated gold[36], which might lead to different conduction channels of our gold structures depending on the temperature and power throughout the applied pulses. An uncertainty of only 25 nm of the current path in the 1 µm wide structure would already suffice to explain the infidelity of our experiments. Measurements on devices with evaporated monocrystalline gold structures[37], which should be less susceptible to this problem, will be the subject of future work.

Even if this limitation cannot be overcome, we foresee several short-term applications for quasi-feedforward decoupling. One direction is Fourier imaging, where the present performance would be sufficient to determine the position of two optically unresolvable nitrogen-vacancy centers with a resolution of 25 nm, comparable to current state-of-the art approaches[24,38]. Another direction are quantum sensing and computation schemes that involve simultaneous manipulation of an electron spin and nuclear qubits[3]. Here, nuclear gates require strong pulses in the MHz regime, and magnetic crosstalk to the electron spin is a source of dephasing. Removing this limitation with our quasi-feedforward scheme could thus expand the range of usable protocols for these experiments.

## Supplementary Material

See supplementary material for details on sample preparation and fabrication, measurement sequence details, the derivation of the estimated single phase gate infidelity, a calculation of the potential imaging resolution of our samples and measurements on the differential probe accuracy.

## Acknowledgements

This work has been supported by the European Union (Horizon 2020 research and innovation programme, grant agreement No 820394 (ASTERIQS)), as well as by the Deutsche Forschungsgemeinschaft (DFG, German Research Foundation) under the German Excellence Strategy – EXC-2111 – 390814868, Emmy Noether grant RE3606/1-1 and the DFG-NFSC program (RE3606/3-1).

## Competing interests

The authors declare that there are no competing interests.

## Author contribution

Georg Braunbeck performed all experiments, was responsible for the diamond preparation, planned the experiment and analyzed the recorded data.

Maximilian Kaindl developed the gold structure, the process to position it around a single NV center and fabricated the structures involved. Andreas Michael Waeber developed the gold structure, the process to position it around a single NV center, fabricated the structures involved and planned the experiment.

Friedemann Reinhard came up with the experiment idea, planned the experiment, analyzed the recorded data and fabricated the structures involved.

Georg Braunbeck and Friedemann Reinhard wrote the paper. All authors commented on the final manuscript.

## Data availability

The data that support the findings of this study are available from the corresponding author upon reasonable request.